\documentstyle[12pt]{article}

\if@twoside
 \else
 \fi

 \parskip \medskipamount
\begin{document}

\hfill{LA-UR-93-2003}

\vspace{7pt}
\begin{center}
{\large\sc{\bf Fermions from photons: Bosonization of  QED in 2+1 dimensions.}}

\baselineskip=12pt
\vspace{35pt}

A. Kovner$^ *$ and P. S. Kurzepa $^{**}$\\
\vspace{10pt}
Theory Division, T-8,
Los Alamos National Laboratory,
MS B-285\\
Los Alamos, NM 87545\\
\vspace{55pt}
\end{center}
\begin{abstract}
We perform the complete bosonization of 2+1
dimensional QED with one fermionic flavor in the
Hamiltonian formalism. The Fermi operators are
explicitly constructed in terms of the vector
potential and the electric field. We carefully specify
the regularization procedure
involved in the definition of these operators, and calculate
the fermionic bilinears and the energy - momentum tensor.
The algebra of bilinears exhibits the Schwinger terms which
also appear in perturbation theory. The bosonic Hamiltonian
is a local, polynomial functional of $A_i$ and $E_i$, and we
check explicitly the Lorentz
invariance of the resulting bosonic theory.
Our construction is conceptually very similar to Mandelstam's
construction in 1+1 dimensions, and is dissimilar from the recent
bosonization attempts
in 2+1 dimensions, which hinge crucially on the presence
of a Chern - Simons term.
\end{abstract}

\vspace{50pt}
\
\newline
*KOVNER@PION.LANL.GOV
\newline
**KURZEPA@MUON.LANL.GOV
\vfill
\pagebreak

\section{Introduction}
It is well known that, in 1+1 dimensions, fermion field
operators can be constructed in terms of local bosonic fields.
This bosonization procedure is of great theoretical
interest in itself, since it provides a method of mutually mapping different
local
quantum field theories with a priori different Hilbert spaces.
In the pioneering
works of Coleman \cite{coleman} and Mandelstam \cite{mandelstam},
the bosonization of a single Dirac field was performed.
Subsequently it was realized, that this procedure is much more general.
An extension was found to theories of several Fermi fields \cite{halpern},
and eventually  non-Abelian bosonization was
introduced by Witten \cite{witten}.
Bosonization has proved to be very helpful in the analysis
of different 1+1 dimensional models. It has been used in a variety of
contexts: relativistic quantum
field theory, condensed matter systems, string theory.

{}From the point of view of particle physics, bosonization techniques have
found perhaps their most interesting applications in the realm of
gauge theories. There they provide insights into the strong coupling
regime of both Abelian
\cite{coleman2}, and non-Abelian \cite{frishman} gauge theories, as well
as better understanding of t'Hooft's solution of large N
$QCD_2$ \cite{affleck}. Also in many other instances, in which
1+1 dimensional theories are used to model the behaviour of $QCD_4$,
bosonization has proven helpful \cite{marek}.

Clearly, the extension of bosonization to higher
dimensions is highly desirable.
It is reasonable to hope that, apart from its intrinsic theoretical interest,
bosonization in higher dimensions will also find diverse applications.
Although it is not likely that bosonized theories in 2+1 and 3+1
dimensions will be exactly solvable, they may be amenable to different
analytical methods than their original fermionic formulations.
A local bosonic formulation of gauge theories with fermions should
also be very interesting from the lattice gauge theory point of view,
since it would provide a convenient starting point for numerical
simulations.

In the late 1970's and early 1980's several attempts,
using different approaches, have been made
to perform bosonization in more
than 1+1 dimensions.
None of them however has lead to a local bosonic theory.
One approach was to perform an exact analog of the Jordan - Wigner
transformation for lattice fermions \cite{susskind}. The resulting bosonic
theory is a $Z_2$ gauge theory, and the bosonic variables are not gauge
invariant, and consequently nonlocal. The nature of the continuum limit of this
bosonic model is also not known. In another approach \cite{luther},
one divides a two (or three) dimensional space into one
dimensional subspaces, and performs one - dimensional bosonization in each
subspace. This procedure, however, does not give local expressions for some
of the local fermionic bilinears. In particular,
the fermion mass term is given by
a complicated nonlocal operator. Related attempts,
using tomografic
projection  \cite{aratyn}, have produced similar results.

After the relative failure of these early approaches,
it has been tacitly assumed untill quite recently,
that the extension of bosonization to higher dimensions is
impossible. The feeling was that 1+1 dimensions is a very special
case, since there is no spin in 1+1 dimensions, and therefore
no ''real'' difference between bosons and fermions.

The renewed interest in this problem was triggered by a
possible connection between the phenomenon of
Fermi - Bose transmutation in 2+1 dimensions, and
novel condensed matter
systems, most notably high $T_c$ superconductors.
Polyakov argued \cite{polyakov}, that the addition of
a Chern - Simons term \cite{deser} to the Lagrangian of
QED$_3$ changes the statistics of charged excitations. His
argument was elaborated further in many works
\cite{kurzepa}, and the explicit realization of this
Chern - Simons induced mechanism of Fermi - Bose
transmutation in lattice theories
has been given in \cite{fradkin}, \cite{luscher}.

There is, however, an important conceptual difference
between the Mandelstam construction in 1+1 dimensions,
and the Chern - Simons bosonization in 2+1 dimensions.
Let us recall the basic bosonization formulae in 1+1
dimensions \cite{mandelstam}. The fermionic operator $\psi_\alpha$
is constructed in terms of a scalar bose field
$\phi$, and its conjugate momentum $p$, as follows,
\begin{equation}
\psi_{1,2}(x)=
\exp{\left [\frac{-2i}{\pi}{\beta}\int_{-\infty}^{x}dyp(y)\pm\frac
{\beta}{2}\phi(x)\right ]}
\end{equation}
which gives the following expressions for the bilinears,
\begin{equation}
\bar\psi\gamma_\mu\psi=
-\frac{\beta}{2\pi}\epsilon_{\mu\nu}\partial_\nu\phi ; \ \
\bar\psi\psi\propto :{\rm cos}\beta\phi:
\end{equation}
One important aspect of these formulae is that the fermion number current,
which in the original theory is conserved due to the equations
of motion, bosonizes
into a topological current, which is conserved trivially. The U(1)
charge, when expressed in terms
of $\phi$, is a topological, rather than a Noether,
charge.
The bosonization is therefore achieved by a duality
transformation: one constructs the fermionic operators
$\psi$ which carry a global U(1) fermion number charge,
in terms of a local bosonic field $\phi$,  which
is itself neutral.

The 2+1 dimensional construction of \cite{fradkin}, \cite{luscher}
is very different in this respect.  (See, however \cite{siva}). In that case
the expression for the Fermi field is \cite{semenoff},
\begin{equation}
\psi(x)=\phi(x)exp{\left [i\int d^2y\theta(x-y)\rho(y)\right ]}
\label{csbos}
\end{equation}
where $\theta(x)$ is the
planar angle, and $\rho$ is the charge density operator.
The fermi field $\psi$
and the bose field $\phi$ carry the same global quantum
numbers. The element of duality is notably missing here.
Moreover, in this picture, a free fermion is bosonized into
a particle interacting with the vector potential with
Chern - Simons action. But the associated bose field is not gauge
invariant, and, after gauge fixing, becomes nonlocal. So one
does not really construct
fermionic operators in terms of {\it local} bosonic fields.

This 2+1 dimensional construction hinges crucially
on the presence of the  Chern - Simons term in the action. The Fermi - Bose
transmutation occurs only for a fixed value of its coefficient.
For other values of this coefficient this procedure gives
nonlocal anyonic fields with fractional statistics.
Another unsatisfactory feature of the procedure is, that it
has been consistently implemented only for nonrelativistic
fermions \cite{fradkin}, \cite{luscher}.
When applied to continuum relativistically invariant
theories \cite{semenoff}, it suffers from regularization ambiguities,
and it is not clear how to interpret the formal results. For example, since
$[\rho(y),\phi(x)]=\phi(x)\delta^2(x-y)$,
and the angle function is not defined at the origin, the ordering on
the right hand side of expression (\ref{csbos}) is completely ambiguous,
and formal manipulations using it are ill- defined.
Furthermore, the
covariant Dirac field has not yet been constructed in this framework,
even formally.

In this paper we take a different approach to bosonization.
Our aim is to construct a Dirac doublet of fermionic operators
in 2+1 dimensions in terms of a {\it local} bosonic field.
The main hypothesis on which we base this construction is the following.
We assume that, in terms of the bose field,
the fermion number charge must be topological,
and thus the bosonized fermion number
current must be trivially conserved \footnote{
In this respect our approach is similar to \cite{marino}, although in this
reference the resulting bosonic theory is nonlocal.}. We are
not aware of any theorem that states this, but this is the case
in all the known examples when either exact bosonization has been
performed (1+1 dimensions) \cite{coleman} \cite{mandelstam}, or
fermionic excitations
are known to exist in the spectrum of a local bosonic theory in
2+1 \cite{ijmp}, or 3+1 \cite{zahed} dimensions.
Another observation which indirectly
supports this assumption is that fermionic fields
are not local fields in the usual sense of the word.
They satisfy local anticommutation, rather than commutation, relations.
If these operators are to be constructed in terms of local bosonic variables,
they must create a nonlocal configuration of the bosonic field. If the
fermion number bosonizes into a topological charge of the bosonic theory,
this will necessarily be true.

Given this assumption, the simplest system that suggests itself as a convenient
object for bosonization is QED with one two-component
Dirac fermion. The reason is that Maxwell's equations,
\begin{equation}
J^\nu=\frac{1}{e}\partial_\mu F^{\mu\nu}
\end{equation}
imply that, if one takes as basic variables electric field $E_i$
and the vector potential $A_i$, the fermion number is trivially
conserved, due to the antisymmetry of $F_{\mu\nu}$. The fermion
number charge is topological,
since it is
equal to the surface integral of $E_i$  {\it at spatial infinity}.
Since there are
no global flavor symmetries in the model,
it is the simplest of its kind.

In this paper we perform the analog of the Mandelstam construction
in the one flavor QED$_3$. The paper is organized as follows. In Section 2
we explicitly construct the two - component
Dirac spinor $\psi_\alpha$  in the Hamiltonian formalism
in terms of bosonic variables $E_i$, and their conjugate
momenta $A_i$. Since the formal definition of $\psi_\alpha$ suffers
from the same kind of ambiguities as in the Chern - Simons case, we
carefully specify the regularization procedure necessary to make it
well - defined.
In physical terms, the fermionic field is an operator
which creates an electric charge $e$ at the point $x$, and also
a pair consisting of a
magnetic vortex and an antivortex of half integer strength $\pi/e$, at
infinitesimal separation.

In Section 3 we use the explicit form of the fermionic operators to calculate
the fermionic bilinears $\bar\psi\Gamma\psi$, and, in particular, show that
the operators $\psi_\alpha$ solve Gauss' constraint.
The calculation is performed
in close analog with the 1+1 dimensional case, by expanding all quantities
in inverse powers of the ultraviolet cutoff.
As in 1+1 dimensions, higher order terms in this expansion
contribute finite renormalization of the coefficients of the lower order
terms. As a result,
Lorentz invariance has to be invoked in order to determine
some coefficients in the expressions for the
spatial components of the current.
We then discuss a certain subtlety encountered in our calculation, and which
is not present in 1+1 dimensions. This problem is due to the presence of
the dimensional
coupling $e$ in 2+1 dimensions, whereas the corresponding couplings
in 1+1 dimensions are dimensionless.
As we show, this restricts the validity of
our bosonization procedure to scales
lower than $\mu$, which is related to  the UV cutoff $\Lambda$ by
$\mu^3=e^2\Lambda^2$.
For finite $e^2$ this is not a real restriction, since the scale $\mu$
becomes infinte in the infinite cutoff limit. However, this restriction
prevents us from
taking the limit $e^2\rightarrow 0$ at finite $\Lambda$.
This we believe, however, is not an artifact of our procedure, but rather a
necessary consequence of the fact that the theory contains only one
two - component fermion. The change of dynamics at scales higher than $\mu$
is necessary to avoid the fermion doubling problem.
We calculate the algebra of the bilinears, and find that it reproduces
the tree level algebra with the addition of Schwinger terms. These
terms are also seen in perturbation theory at the one loop level.

In Section 4 we find the bosonized expression for the
energy - momentum tensor. Again, as with the spatial components of
the current,
we use Lorentz invariance to determine several coefficients. It is shown that,
with the right choice of these coefficients,
the resulting theory is Lorentz -
 invariant in the limit $\Lambda\rightarrow\infty$, and one recovers
Maxwell's equations as its equations of motion.

Section 5 is devoted to the discussion, and some extensions of our results.
In particular,
we indicate how the well known regularization ambiquity of one
flavor QED$_3$ \cite{coste} appears in our framework. It is possible
to construct a modified
Fermi field $\chi_\alpha$, which solves the modified
Gauss' constraint $\chi^{\dagger}\chi=1/e\partial_iE_i+en/2\pi B$.
The coefficient $n$ must be an integer, to ensure the correct quantization
of the fermion number charge. The new bosonized theory differs from the old
one precisely by an addition of a Chern-Simons term
to the Hamiltonian \cite{inprep}.

In Appendices we discuss the simple intuitive mechanism underlying
the anticommutation
relations of $\psi$, discuss the geometrical aspects of our construction,
and give some details of the calculation of
the bilinears.

\section{The construction of Fermi operators}
The 2+1 dimensional quantum electrodynamics is defined by the following
 Hamiltonian (we work in the timelike, or Weyl,	gauge $A_0=0$),
\begin{equation}
H=\frac{1}{2}E^2+
\frac{1}{2}B^2+\bar\psi\gamma^i(i\partial_i+eA_i)\psi+m\bar\psi\psi
\label{ham}
\end{equation}
together with the Gauss' constraint,
\begin{equation}
\partial_iE_i=e\psi^\dagger\psi
\label{const}
\end{equation}
The two component Fermi field $\psi_\alpha$, $\alpha=1,2$, and the bosonic
variables $E_i$, and $A_i$ satisfy the canonical (anti)commutation relations,
\begin{equation}
\{\bar\psi_\alpha (x),\psi_\beta (y)\}=\delta_{\alpha\beta}\delta^2(x-y);\ \
[E^i(x),A^j(y)]=i\delta^{ij}\delta^2(x-y)
\end{equation}
As is usually the case with gauge theories, the Hamiltonian eq.(\ref{ham})
acts on a large Hilbert space, which contains unphysical degrees of
freedom.
Those must be ultimately eliminated by solving
the constraint eq.(\ref{const}).
One usually solves the constraint by expressing the longitudinal part of
the electric
field in terms of the matter fields.
The complete set of operators that are then used to span
the physical Hilbert space consists of the transverse part of
$E_i$, and the matter fields $\psi$.
In the context of our problem, however, we view eq.(\ref{const}) as
one of the bosonization equations.
We will therefore retain both components of the electric field, and instead
solve the Gauss' constraint by constructing
the doublet of anticommuting fermionic operators $\psi_\alpha$
in terms of $E_i$, and $A_i$ \footnote{Since the fermion field
in QED$_3$
carries only one global quantum number, namely, the electric charge, it
should be representable in terms of one bosonic field. This bosonic field
will not be free of course, but will rather be strongly interacting.
Our construction gives precisely this counting: the fields $E_i$ and their
conjugate momenta describe a photon, and an additional
scalar degree of freedom.}.

Substituting those back into the Hamiltonian
eq.(\ref{ham}), we will obtain the completely bosonized form of
the theory, defined on the physical Hilbert space of QED$_3$.
Our bosonization procedure is therefore defined only for {\it gauge invariant}
quantities.
We will also calculate the fermionic bilinears, and find that
they satisfy the tree level algebra, modified by Schwinger terms.
The appearance of these Schwinger terms is also
seen in perturbation theory at the one loop level.

To construct the operators $\psi_\alpha$, we must first fix the gauge freedom
associated with the time -  independent gauge transformations, generated
by the Gauss' constraint. We do this by considering $\psi$ in the
Coulomb gauge. Those are the gauge - invariant operators,
\begin{equation}
\psi^{CG}_\alpha(x)=\psi_{\alpha}(x)
exp{\left [ie\int d^2y e_i(y - x)A_i(y)\right ]}
\label{cog}
\end{equation}
where
\begin{equation}
e_i(x)=-\frac{1}{2\pi}\frac{x_i}{x^2}
\end{equation}
is the electric field of a point charge.
The exponential factor ensures the gauge invariance of the operator
$\psi^{CG}$. In the following we will omit the superscript $CG$, and always
understand that the fermionic operators we are constructing are defined by
eq.(\ref{cog}).
It is important to note, that the Hamiltonian eq.(\ref{ham}) can be
rewritten in terms of these operators.  In this formulation, the
covariant derivative in eq.(\ref{ham})
contains only the transverse part of the vector potential $A_i$, $A_i^T$.

In addition to solving the constraint eq.(\ref{const}), the
fermionic operators $\psi_\alpha$ must satisfy the following
conditions \footnote{Bosonization of the massive Schwinger model
in 1+1 dimensions can be formulated in precisely
the same terms.}:

i. Carry unit electric charge,
$[\psi_\alpha(x),\partial_iE_i(y)]=e\psi_\alpha(x)\delta^2(x-y)$;

ii. Transform correctly under rotations. For convenience, we
choose the basis of Dirac matrices in which the rotation
generator is diagonal,
\begin{equation}
\gamma^0=\sigma^3 ;\  \gamma^1=i\sigma^2 ;\ \ \gamma^2=-i\sigma^1
\end{equation}
Then,
\begin{equation}
\psi_1\rightarrow e^{i\phi/2}\psi_1 ; \ \ \psi_2\rightarrow e^{-i\phi/2}\psi_2
\label{rot}
\end{equation}
 where $\phi$ is the rotation angle;

iii. Fermionic bilinears must be local operators,
$[\psi^\dagger_\alpha\psi_\beta(x),O(y) ]=0$ for $x\ne y$, and
for any local gauge invariant operator $O(x)$.

To satisfy the first condition, we take the following ansatz,
\begin{equation}
\psi_\alpha(x)=k\Lambda V_\alpha(x)\Phi(x)U_\alpha(x)
\label{an1}
\end{equation}
where,
\begin{equation}
\Phi(x)=exp{\left [ie\int d^2y e_i(y - x)A_i(y)\right ]}
\label{P}
\end{equation}
Here $\Lambda$
is the ultraviolet cutoff, $k$ is a finite dimensionless constant,
which depends on the precise definition of the UV cutoff, and the
operators $V_\alpha$
and $U_\alpha$ commute with the charge density operator.
To ensure the anticommutativity
of $\psi$'s, we take the following forms for $U_\alpha$ and $V_\alpha$,
\begin{eqnarray}
V_1(x)=-iexp{\left [\frac{i}{2e}\int d^2y(\theta(x-y)-
\pi)\partial_iE_i(y)\right ]}&;& V_2(x)=-i V^\dagger_1(x); \\ \nonumber
U_1(x)=exp{\left [-\frac{i}{2e}\int d^2y\theta(y-x)\partial_iE_i(y)\right ]}&;&
U_2(x)=U^\dagger_1(x)
\label{uv1}
\end{eqnarray}
where $\theta(x)$ is the polar angle at the
point $x$.

With these definitions we find,
\begin{eqnarray}
\psi_1(x)\psi_1(y)&=&\psi_1(y)\psi_1(x)
e^{-i(\theta(y-x)-\theta(x-y))}=-\psi_1(y)\psi_1(x) \\ \nonumber
\psi_1(x)\psi^\dagger_1(y)&=&\psi^\dagger_1(y)\psi_1(x)
e^{i(\theta(y-x)-\theta(x-y))}=-\psi^\dagger_1(y)\psi_1(x); \\ \nonumber
\psi_2(x)\psi_2(y)&=&\psi_2(y)\psi_2(x)e^{i(\theta(y-x)-\theta(x-y))}=
-\psi_2(y)\psi_2(x) \\ \nonumber
\psi_2(x)\psi^\dagger_2(y)&=&\psi^\dagger_2(y)\psi_2(x)
e^{-i(\theta(y-x)-\theta(x-y))}=-\psi^\dagger_2(y)\psi_2(x); \\ \nonumber
\psi_1(x)\psi^\dagger_2(y)&=&\psi^\dagger_2(y)\psi_1(x)e^{-i\pi}=
-\psi^\dagger_2(y)\psi_1(x) \\ \nonumber
\psi_1(x)\psi_2(y)&=&\psi_2(y)\psi_1(x)e^{i\pi}=-\psi_2(y)\psi_1(x)
\end{eqnarray}
The factor $k\Lambda$ in eq.(\ref{an1}) ensures the
correct dimensionality of the fermionic fields, and
the correct normalization of the anticommutators at coincident points,
\begin{equation}
\{\psi^\dagger_1(x),\psi_1(y)\}=\{\psi^\dagger_2(x),\psi_2(y)\}=\delta^2(x-y)
\end{equation}

Here a comment is in order about the precise meaning of
eqs.(\ref{an1}-\ref{uv1}).
The function $\theta(x)$ is defined only modulo $2\pi$. On the other hand,
the eigenvalues of the charge density operator are quantized in
units of $1/e$.  As a consequence, the operators
$U_\alpha$ and $V_\alpha$ (and thus $\psi_\alpha$) are double - valued.
This, in fact, is to be expected on the general grounds.
Under rotation by $2\pi$
the fermionic operators change sign, but the bosonic operators $A_i$ and $E_i$
are unchanged. Therefore $\psi$ cannot be a single valued function of
$A_i$, $E_i$, and the coordinate $x$ \cite{finkelstein}. Further
geometrical details of the construction, and its interpretation along the
lines of \cite{finkelstein} are contained in the Appendix B.

These manipulations are, however, too formal, since the commutators of $V(x)$
and $U(x)$ with $\Phi(x)$ in eq.(\ref{an1}) are singular. The singularity
is due to the fact that $\Phi(x)$ creates
a charge at the point $x$, and consequently the commutator involves the
ill - defined  factor
$e^{i\theta(0)/2}$.
The expression for the fermionic operators needs
regularization. The natural way to regularize it is by point splitting.
We will therefore consider the following expressions,
\begin{equation}
\psi^\eta_1(x)=
V_1(x+\eta)\Phi(x)U_1(x-\eta);
\label{psieta}
\end{equation}
$$\psi^\eta_2(x)= V_2(x+\eta)\Phi(x)U_2(x-\eta)$$
The length of the regulator $\eta$ is taken to be proportional to the
inverse of the UV cutoff $|\eta|\propto 1/\Lambda$.
However, considering   $\eta$ in any fixed direction breaks the rotational
symmetry of the problem. To restore rotational invariance, we average over
the direction of $\eta$. The averaging
should be performed with an appropriate phase factor, in order to ensure
the correct rotational properties of the operators.
Under rotation by an angle $\phi$, $\theta$ transforms
as follows, $\theta(x)\rightarrow\theta(x)+\phi$.
We arrive therefore at,
\begin{equation}
\psi_1(x)=\lim_{\Lambda\rightarrow \infty}\frac{k\Lambda}{2\pi }
\int d\hat\eta e^{-i\frac{\theta(\eta)}{2}}\psi_1^{\eta}(x), \ \
\psi_2(x)=\lim_{\Lambda\rightarrow \infty}\frac{k\Lambda}{2\pi }
\int d\hat\eta e^{i\frac{\theta(\eta)}{2}}\psi_2^{\eta}(x)
\label{psi}
\end{equation}
where the integral is
over the angle of the vector $\eta$, and $\hat\eta$ is the unit vector
parallel to $\eta$.
It can be explicitly checked, that,
\begin{equation}
\{\psi_\alpha(x),\psi_\beta(y)\}=0, \ \ |x-y|>>1/\Lambda
\end{equation}
Therefore, in the limit $\Lambda\rightarrow \infty$, we regain the
standard anticommutation relations.

However, we are still not done with the construction of $\psi$. The point
that remains to be settled is the following. As defined up to now,
the fermionic operators depend only on the longitudinal
component of the electric
field $E^L_i$. Any bilinear that one would calculate using these
expressions would also depend only on $E^L_i$, and not on the
transverse part $E^T_i$. However, $E^L_i$ is not a local field, and it
is therefore impossible to  obtain local bilinears with this definition of
fermionic operators.
(We have checked this by explicit calculation.)
It turns out, that this problem can be remedied by modifiyng
the expressions for $V(x)$ and $U(x)$, so
that they become
creation and annihilation operators of a magnetic vortex of
half - integer strength,
\begin{equation}
V_1(x)=-i\exp\left\{\frac{i}{2e}\int
d^2y\left[(\theta(x-y)-\pi)\partial_iE_i(y)
+2\pi G^{(2)}(y-x)\epsilon_{ij}\partial_iE_j(y)\right]\right\};
\label{vu}
\end{equation}
$$U_1(x)=
\exp\left\{-\frac{i}{2e}\int d^2y\left[\theta(y-x)\partial_iE_i(y)+2\pi
 G^{(2)}(y-x)\epsilon_{ij}\partial_iE_j(y)\right]\right\}; $$
$$V_2(x)=-iV^\dagger_1(x)  ;  U_2(x)=U^\dagger_1(x)$$
Here $G^{(2)}(x-y)=-\frac{1}{4\pi}\ln(\mu^2 x^2)$, $x^2=x^ix^i$, is the
Green's function of the two - dimensional Laplacian, with IR cutoff
$\mu$.

The physical meaning of the operators $V_\alpha$ and $U_\alpha$ is
clear from the following commutation relations \cite{npb},
\begin{equation}
[V_1(x),B(y)]=-\frac{\pi}{e}V_1(x)\delta^2(x-y)  ;\ \
[U_1(x),B(y)]=\frac{\pi}{e}U_1(x)\delta^2(x-y)
\end{equation}
This modification does not change either the anticommutation
relations or the rotational properties of $\psi$.
Moreover, now the point splitting procedure used
in regularizing $\psi$ has a
very natural interpretation, since
the vortex operators $V$ and $U$ are
the dual variables of $QED_3$ \cite{marino1}. In any lattice
regularized version of the theory the charged
field $\Phi$ should live on the lattice sites,
and the vortex fields should live on the sites of
the dual lattice, and therefore be point split from $\Phi$.

Equations (\ref{vu}), (\ref{psi}) and (\ref{psieta}) are our
final expressions for
the fermionic operators in terms of the bose fields $E_i$ and $A_i$.

\section{Calculation of bilinears}

Our next step is to calculate fermionic bilinears which do not contain
derivatives.
As usual, those should
be defined with the help of
a point - splitting procedure. We use the following
definition,
\begin{equation}
J_\Gamma(x)=\bar\psi(x)\Gamma\psi(x)\equiv \frac{1}{8\pi}
\int d\hat\epsilon e^{i\chi_\Gamma(\hat\epsilon)}\left\{\left[
\psi^\dagger(x+\epsilon),\gamma^0\Gamma\psi(x-\epsilon)\right]
,e^{ie\int_{x-\epsilon}^{x+\epsilon}dx_i
A_i^{T}}\right\}_{|\epsilon|,|\eta|\propto 1/\Lambda}
\label{bilinear}
\end{equation}
It is implicitly understood in eq.(\ref{bilinear}) that the
limit $\Lambda\rightarrow \infty$ is taken at the end, after
(independent) averaging over the directions of $\epsilon$ and $\eta$.
The insertion of Wilson factor is appropriate for the definition
of bilinears in a gauge theory. Since we are constructing the
Coulomb gauge fermions, it is only the transverse part of the vector
potential that appears in the Wilson factor. The phase
$\exp\{i\chi_\Gamma\}$ is inserted to project onto
the relevant irreducible representation of the 2D
rotation group, while averaging over $\hat \epsilon$.
Thus, for $\Gamma=\gamma_0$ and $\Gamma=1$ we have
$\chi(\epsilon)=0$, while for $\Gamma=\gamma_+\equiv\gamma_1+i\gamma_2$,
$\chi(\epsilon)=\theta(\hat\epsilon)$, and for $\Gamma=\gamma_
-\equiv\gamma_1-i\gamma_2$, $\chi(\epsilon)=-\theta(\hat\epsilon)$.
The ratio $\eta/\epsilon$ is arbitrary, but the final results should not
depend on it.

The calculation of bilinears proceeds
in complete analogy to the 1+1 dimensional case. Namely, we expand
the expression eq.(\ref{bilinear}) in powers of the inverse
cutoff, and retain only the terms that do not vanish in the continuum limit
$\Lambda\rightarrow\infty$.
This procedure has a certain caveat that one has to keep in mind.
The operators which are multiplied by inverse powers of $\Lambda$, and
are formally small, may, in fact, give finite contribution in the continuum
limit, if these operators have  high enough dimensions. This problem, in fact,
arises also in 1+1 dimensions, where the effect of the higher order terms is
to renormalize the coefficients of the lower dimensional operators in a way
consistent with the symmetries of the problem \cite{mandelstam}. We will return
to this point later, and argue that the situation in our case is
similar.

Let us begin by calculating the charge density and the mass operators.
For that we need $\psi^\dagger_1\psi_1$, and $\psi^\dagger_2\psi_2$,
\begin{equation}
\psi^\dagger_{1\xi}(\epsilon)\psi_{1\eta}(-\epsilon)=
e^{-\frac{i}{2}[\theta(\eta)-\theta(\xi)+\frac{1}{2}R_1(\epsilon,\xi,\eta)]}
e^{ie\int d^2yf_i(y)A_i(y)+
\frac{i}{2e}\int d^2ya^1_L(y)\partial_iE_i-
a^1_T(y)\epsilon_{ij}\partial_iE_j(y)}
\label{11}
\end{equation}
where the c - number phase $R_1$ is given by,
\begin{equation}
R_1(\epsilon,\xi,\eta)=\theta(\xi-2\epsilon)
-\theta(\xi+2\epsilon)+\theta(\eta-2\epsilon)-\theta(\eta+2\epsilon)
\end{equation}
and the functions appearing in eq.(\ref{11}) are defined as,
\begin{equation}
f_i(y)\equiv e_i(y+\epsilon)-e_i(y-\epsilon);
\end{equation}
$$
a^1_L(y)\equiv \theta(y-\epsilon+\xi)
-\theta(\epsilon+\xi-y)+\theta(\eta-\epsilon-y)-\theta(y+\epsilon+\eta)
$$
$$
\frac{1}{2\pi}
a^1_T(y)\equiv G^{(2)}(y-\epsilon+\xi)
-G^{(2)}(\epsilon+\xi-y)+G^{(2)}(\eta-\epsilon-y)-G^{(2)}(y+\epsilon+\eta)
$$

We now have to expand the operatorial part in the exponential
in eq.(\ref{11}) in Taylor series in $\epsilon_i$, $\xi_i$, and $\eta_i$.
The only subtlety here is that the derivatives do not commute
when acting on $\theta(x)$, and thus the order of derivatives
has to be specified.
Keeping in mind the physical picture, however, it is  clear that
we must first expand all expressions in powers of
$\eta_i$ and $\xi_i$ at fixed $\epsilon_i$, and only afterwards expand in
powers of $\epsilon$. The order of taking the derivatives
is therefore unambigous. Expanding eq.(\ref{11}) up to terms of
the second order, we obtain (for details see Appendix C),
\begin{equation}
J_{11}\equiv\psi^\dagger_1\psi_1=
2\pi k^2\Lambda^2\left[\frac{i}{e}<(\eta+\xi)_i\epsilon_j>\partial_j\tilde E_i-
<\epsilon_i(\eta_j-\xi_j)>\{A_i,\tilde E_j\}\right]
\end{equation}
Here $\tilde E_i\equiv\epsilon_{ij}E_j$, and the averaging
is defined as follows,
\begin{equation}
<\alpha>\equiv
\frac
{i}{8\pi^3}\int d\hat\epsilon d\hat\eta d\hat\xi \alpha
e^{-i[\theta(\eta)-\theta(\xi)]}{\rm Im}
e^{-\frac{i}{4}R_1(\epsilon,\xi,\eta)}
\end{equation}
Performing the averaging, and choosing the constant $k$  appropriately, we get,
\begin{equation}
\psi^\dagger_1\psi_1=\frac{1}{2e}\partial_i E_i-A\cdot\tilde E
\label{11a}
\end{equation}
The calculation of $\psi^\dagger_2\psi_2$ proceeds analogously,
and gives the parity conjugate of eq.(\ref{11a}),
\begin{equation}
\psi^\dagger_2\psi_2=\frac{1}{2e}\partial_i E_i+A\cdot\tilde E
\end{equation}
We obtain therefore for the charge density $J_{0}$ and the mass term $J$,
\begin{equation}
J_0\equiv\psi^\dagger\psi=\frac{1}{e}\partial_iE_i
\label{charge}
\end{equation}
\begin{equation}
J\equiv\bar\psi\psi=-2A\cdot\tilde E
\label{mass}
\end{equation}
The correction to these expressions are of the order $1/\Lambda^2$, and we
neglect them.

We now turn to the calculation of the spatial
components of the current. Following the same steps as in the
previous derivation, we find,
\begin{equation}
\psi^\dagger_{2\xi}(\epsilon)\psi_{1\eta}(-\epsilon)=-
e^{\frac{i}{2}[-\theta(\eta)-\theta(\xi)+\frac{1}{2}
R_2(\epsilon,\xi,\eta)]}e^{-i\pi\int d^2y \frac{1}{e}\partial_iE_i(y)}
\label{12}
\end{equation}
$$e^{ie\int d^2yf_i(y)A_i(y)-
\frac{i}{2e}\int d^2ya^2_L(y)\partial_iE_i+a^2_T(y)
\epsilon_{ij}\partial_iE_j(y)}
$$
The phase $R_2$ is given by,
\begin{equation}
R_2(\epsilon,\xi,\eta)=\theta(\xi-
2\epsilon)-\theta(\xi+2\epsilon)-\theta(\eta-2\epsilon)+\theta(\eta+2\epsilon)
\end{equation}
and,
\begin{equation}
a^2_L(y)\equiv \theta(y-\epsilon+\xi)-\theta(\epsilon+\xi-y)-
\theta(\eta-\epsilon-y)+\theta(y+\epsilon+\eta)
\end{equation}
$$
\frac{1}{2\pi}
a^2_T(y)\equiv G^{(2)}(y-\epsilon+\xi)
-G^{(2)}(\epsilon+\xi-y)-G^{(2)}(\eta-\epsilon-y)+
G^{(2)}(y+\epsilon+\eta)
$$
This again is to be expanded in powers of the inverse cutoff. After some
algebra (see Appendix C), and keeping two terms in the expansion, we obtain,
\begin{equation}
J_{21}=
-k^2\Lambda^2
2<\epsilon_i>eA_i-
\label{j12}
\end{equation}
$$
-
k^2\Lambda^2
\left[
\frac{e}{3}
<\epsilon_i\epsilon_j\epsilon_k>
\partial_i\partial_jA_k-\frac{4e^3}{3}<\epsilon_i\epsilon_j\epsilon_k>
A_iA_jA_k\right]
$$
$$-4\pi^2k^2\Lambda^2
\left[\frac{1}{e}
<\epsilon_i(\xi_j+\eta_j)(\xi_k+\eta_k)>A_i\tilde E_j\tilde E_k
+\frac{i}{e^2}
<\epsilon_i(\xi_j-\eta_j)(\xi_k+\eta_k)>\partial_i\tilde E_j\tilde E_k\right]
$$
where the averages are now defined as,
\begin{equation}
<\alpha>\equiv\frac
{i}{8\pi^3}\int d\hat\epsilon d\hat\eta d\hat\xi \alpha
e^{-i[\theta(\eta)+\theta(\xi)+
\theta(\epsilon)]}{\rm Re}
e^{\frac{i}{4}R_2(\epsilon,\xi,\eta)}
\end{equation}
Already at this stage we see that the calculation
of the spatial components of the current is more
involved. The leading term is proportional to the UV cutoff.
The next - to - leading term in the
expansion is of order $1/\Lambda$, but its commutator
with the leading term is finite. We
therefore have to keep at least those next order terms which give a
contribution to this commutator, even though they vanish in
the naive continuum limit.

We thus have to come to grips here with the problem mentioned at the
beginning of this section, namely, that discarding the formally
small higher order
terms in the above expansions is too naive. The simplest manifestation of this
problem is the following.
Explicitly evaluating the averages in eq.(\ref{j12}),
and keeping only the leading terms, one obtains,
\begin{equation}
J_i\equiv\bar\psi\gamma_i\psi=-e\kappa \Lambda A_i+
\frac{1}{e\Lambda}\left[
\beta\tilde E_i(A\tilde E)+\gamma E^2A_i\right]
\label{current}
\end{equation}
However, the constants $\kappa$, $\beta$ and $\gamma$ in the above formula
depend on the ratio of the regulators $|\eta|/|\epsilon|$.
On the other hand, it is clear that the final result may
not depend on this ratio. This
means, that the formally small terms that we discarded do
not disappear completely, but renormalize the constants in
eq.(\ref{current}).
The situation
here is precisely analogous to the 1+1 dimensional case
\cite{mandelstam}. There, the same naive expansion procedure
gave an incorrect overall scale of $j_1$, which resulted in  $j_0$ and $j_1$
not forming  a Lorentz vector.
In 1+1 dimensions
one could, in principle, explicitly evaluate
all the corrections coming from the higher order terms.
Instead, the requirement of Lorentz invariance was
sufficient to fix the scale of $j_1$.
In the present case we use precisely the same method. In the next section,
we will calculate the energy - momentum tensor,
and require that the theory be Lorentz invariant in the continuum limit
$\Lambda\rightarrow\infty$.
We will also require that the spatial components of the current,
together with the mass term, satisfy (up to possible Schwinger terms)
the tree - level algebra.

Now it remains to convince ourselves, that the
contribution of the higher order
terms leads to {\it finite} renormalization of the coefficients in the
expressions for the $J_i$'s. In 1+1
dimensions this is a trivial consequence of the fact that the coupling
constant in, say the sine - Gordon theory, is dimensionless.
Since the coupling
constant $e$ in QED$_3$ is dimensionfull, we cannot use similar arguments here.
In fact, at first glance, it seems that
some of the terms we have discarded are actually more important than the ones
we have kept. The expansion in powers of $1/\Lambda$ will bring
down terms of the same general form as in eq.(\ref{current}), but
multiplied by powers of the factors
of the following three types,
\begin{equation}
\frac{\partial_i}{\Lambda}; \ \ \frac{eA_i}{\Lambda}; \ \ \frac{E_i}{e\Lambda}
\end{equation}
The first type of factor is harmless, since, clearly,
it can lead only to a finite
renormalization of lower -  dimensional terms.
Since the scaling dimension of the
vector potential in perturbation theory is $1/2$, the second factor
is small at any physical scale, and can be discarded. The third term,
however, looks dangerous. The perturbative dimension of $E_i$ is $3/2$,
and this factor seems therefore to be of order $\Lambda^{1/2}$. If that were
true, it would imply that the terms containing higher powers of this factor
are more important than the lower order terms\footnote{
Although in QED$_2$ the coupling constant has the dimension of mass,
the electric field does not have a transverse component. Therefore
the scaling dimension of $E/e$ is zero, and the problem does not arise.}.

This conclusion
is not correct, however, for the following, rather subtle, reason.
It turns out that, apart from the UV cutoff $\Lambda$, the bosonized
theory has an additional UV scale $\mu=(e^2\Lambda^2)^{1/3}$.
The appearance of this scale can be seen as follows.
The fermionic operator defined in eq.(\ref{psi}) is essentially
a product of a vortex and
an antivortex operator, separated by $\Lambda^{-1}$.
The UV scaling behavior of the vortex
operator in perturbation theory is  known \cite{npb}. At short distances
it scales as an exponential,
\begin{equation}
<V(x)V^*(y)>\propto|x-y|^\alpha
exp{\left [\frac{c}{e^2}|x-y|^{2 - \alpha }\right ]}
\end{equation}
where $c$ is a constant, and $\alpha /2$ is the scaling dimension of the
electric field, $<E(x)E(y)>\propto \frac{1}{|x-y|^\alpha }$.
Assuming the perturbative scaling of the electric field
at short distances, $\alpha = 3$,  we find that
the fermion operator that we have constructed, at distances
larger than $1/\Lambda$, but still small relative to any physical distance
scale, behaves, {\it roughly}, as,
\begin{equation}
<\psi^\dagger_\eta(x)\psi_\xi(y)>\propto|x-y|^\gamma
exp{\left [\frac{c\hat\eta\hat\xi}{e^2\Lambda^2
|x-y|^3}\right ]}
\label{psicor}
\end{equation}
The scale $\mu$ is therefore clearly a {\it crossover scale}.
At distances larger than,
\begin{equation}
|x-y|^3\propto \frac{1}{e^2\Lambda^2}\equiv\frac{1}{\mu^3}
\label{dist}
\end{equation}
the exponential in eq.(\ref{psicor}) can be expanded in power series.
In that case
the fermion propagator scales as a power, which is consistent
with its perturbative behavior.

At distances smaller than $1/\mu$
the non - point - likeness of $\psi$ becomes important.
As a result the  scaling behavior at these short distances
in the bosonized theory must be different
from the perturbative one.
In fact, if the fermionic operator is still to scale with a power law
for $\mu<<1/x<<\Lambda$, the leading UV behavior of the propagator of
the electric field must be $<E_i(x)E_i(y)>_{|x-y|<1/\mu}
\sim \frac{e^2}{|x-y|^2}$.
We will return to this point later, and see that this short distance
behavior emerges
naturally from the bosonic Hamiltonian.

Assuming for the moment that this is
indeed the correct asymptotics, we find that our procedure of calculation
of the bilinears is indeed self - consistent.
The order of magnitude of the "dangerous" correction factor is,
\begin{equation}
\frac{E_i}{e\Lambda}\propto\frac{\mu^{3/2}}{e\Lambda}=O(1)
\end{equation}
and the corrections due to the higher order
terms in the expansion of the bilinears can only
lead to a finite renormalization of the lower order terms.

We therefore take the current in the general form,
\begin{equation}
J_i\equiv\bar\psi\gamma_i\psi=-e\kappa \Lambda A_i+
\frac{1}{e\Lambda}\left[
:\beta\tilde E_i(A\tilde E)+\gamma E^2A_i:\right]
\label{urrent}
\end{equation}
Clearly, to make this expression well defined,
we must normal order it with respect to the perturbative vacuum.
This implies the subtraction of all terms with at least one
Wick contracted pair of operators. It has been rigorously shown that,
in 1+1 and 2+1 dimensions, Wick ordering with respect to the free
measure of fields in an {\it interacting} theory always removes
the most divergent parts of the correlation functions \cite{glimm}.  We assume
that, in the present case,
this is sufficient to subtract the
divergent part of the operator in the square brackets in eq.(\ref{urrent}).

To be more specific, we {\it assume} the following form of the UV asymptotics
of the correlators of the bosonic fields, which is the most general one
consistent with the UV scaling dimensions discussed above, rotational symmetry,
and parity transformation properties,
\begin{eqnarray}
lim_{x\rightarrow y}<A_i(x)A_j(y)>&=&r_1\frac{\delta_{ij}}
{|x-y|}+2r_2\frac{(x-y)_i(x-y)_j}{|x-y|^3}; \\ \nonumber
lim_{x\rightarrow y}\frac{1}{e^2}<E_i(x)E_j(y)>&=&q_1
\frac{\delta_{ij}}
{|x-y|^2}+2q_2\frac{(x-y)_i(x-y)_j}{|x-y|^4}; \\ \nonumber
lim_{x\rightarrow y}<E_i(x)A_j(y)>&=&s_1\frac
{\delta_{ij}}{|x-y|^2}+2s_2\frac{(x-y)_i(x-y)_j}{|x-y|^4}\\ \nonumber
&+&mp_1\frac{\epsilon_{ij}}{|x-y|}+2mp_2\frac
{(\tilde x-\tilde y)_i(x-y)_j-
(\tilde x-\tilde y)_j(x-y)_i}{|x-y|^3}; \\ \nonumber
\label{correl}
\end{eqnarray}
The expectation values of the quadratic operators are then given by,
\begin{equation}
\frac{1}{e^2}<E^2>=2(q_1+q_2)\Lambda^2; \ \  <A^2>=2(r_1+r_2)\Lambda
\label{expv}
\end{equation}
$$
<A\cdot E>=2(s_1+s_2)\Lambda^2; \ \ <A\cdot\tilde E>=-2m(p_1+p_2)\Lambda
$$
$$ <B^2>=-(r_1+2r_2)\Lambda^3$$

The numerical coefficients are therefore subject to the following conditions,
\begin{equation}
 q_1+q_2>0; \ \ r_1+r_2>0; \ \ r_1+2r_2<0; \ \ s_1+s_2=0
\label{coef}
\end{equation}
The first three conditions are obvious, while the last one should hold in a
Lorentz invariant theory. It follows from the fact that $E\cdot J$ is the
zeroth component of a Lorentz vector, and therefore must
have a vanishing expectation value.

In order to fix the coefficients in eq.(\ref{urrent}),
we first impose the condition that,
together with other {\it normal ordered} bilinears, this current
satisfies the tree - level current algebra,
up to possible Schwinger terms, and terms that vanish in the continuum limit,
\begin{equation}
[J_i(x),J_j(y)]=2i\epsilon_{ij}J(x)\delta^2(x-y) ;\ \
[J_i(x),J(y)]=-2i\epsilon_{ij}J_j(x)\delta^2(x-y)
\label{su2}
\end{equation}
Evidently, with the mass bilinear given by eq.(\ref{mass}),
the second equation holds without any corrections. Calculating the first
commutator, we obtain,
\begin{equation}
[J_i(x),J_j(y)]=i\epsilon_{ij}\{-2(q_1+q_2)(p_1+p_2)(\beta+2\gamma)^2
+[-\kappa(3\beta+2\gamma)+(q_1+q_2)(\beta+2\gamma)^2]:A\cdot\tilde E:
\end{equation}
$$
+\frac{\beta+2\gamma}{e^2\Lambda}[-m(p_1+p_2)
(\beta+2\gamma):E^2:+\frac{\beta+\gamma}{\Lambda}:E^2A\cdot\tilde E:]\}
$$
This gives two conditions,
\begin{equation}
\beta+2\gamma=0; \ \ \beta=\frac{2}{\kappa}
\end{equation}

The current is then determined as,
\begin{equation}
J_i\equiv\bar\psi\gamma_i\psi= - e\kappa\Lambda A_i+
\frac{1}{e\kappa\Lambda}\left[2\tilde E_i(A \tilde E)
- E^2A_i +4(p_1+p_2)m\Lambda\tilde E_i\right]
\label{Current}
\end{equation}
The constant $\kappa$ will be determined in the following section.

The fermionic bilinears are now expressed as local
functions of $E_i$ and $A_i$. Their algebra can be calculated explicitly,
\begin{equation}
[J_i(x),J_j(y)]=2i\epsilon_{ij}J(x)\delta^2(x-y) ;\ \
[J_i(x),J(y)]=-2i\epsilon_{ij}J_j(x)\delta^2(x-y)
\label{algebra}
\end{equation}
$$[J_0(x),J_i(y)]=-i\left[\kappa\Lambda\delta_{ij}+
\frac{2}{e^2\kappa\Lambda}(E_iE_j-1/2E^2\delta_{ij})
\right]\partial^x_j \delta^2(x-y);$$
$$[J(x),J_0(y)]=\frac{2i}{e}E_i(x)\epsilon_{ij}\partial_j^x\delta^2(x-y)
$$
The first two commutators are the canonical tree - level ones.
The other two exhibit explicitly the Schwinger terms mentioned above.
This, again, is a common feature of our procedure and the bosonization in 1+1
dimensions: the Schwinger terms, which in perturbation theory appear at the
one loop level, appear  in the bosonized theory in the tree level commutators.

\section{The Hamiltonian and Lorentz invariance}

We now calculate the energy - momentum tensor of the theory
in terms of bosonic fields. The gauge invariant, symmetric
energy-momentum tensor of QED$_3$ is given by,
\begin{equation}
T^{\mu\nu}=T^{\mu\nu}_B+T^{\mu\nu}_ F
\end{equation}
with the bosonic and fermionic parts,
\begin{equation}
T^{\mu\nu}_B=
F^{\mu\lambda}F_{\lambda}^\nu+\frac{1}{4}g^{\mu\nu} F^2
\end{equation}
\begin{equation}
T^{\mu\nu}_F=\frac{i}{4}(\bar\psi\gamma^\mu D^\nu\psi+\bar
\psi\gamma^\nu D^\mu\psi-D^\nu\bar\psi\gamma^\mu\psi
-D^\mu\bar\psi\gamma^\nu\psi)
\label{enmom}
\end{equation}
Here $D=\partial - ieA$ is the covariant derivative.
In order to bosonize the fermionic part of the energy - momentum tensor,
one has to calculate the bilinears in eq.(\ref{enmom}).
Their direct calculation in terms of the fermionic operators
constructed in Section 2 gives the general form of the different
terms which enter the Hamiltonian,
but cannot establish their coefficients (because of the
unknown finite renormalization due to the higher order terms).

In 1+1 dimensions there is an alternative way of calculating $T^{\mu\nu}$,
using its representation in the Sugawara
form. The energy - momentum
tensor in a free fermionic theory can be written as a suitably regularized
product
of the currents \cite{avatars}.
In 1+1 dimensions it is,
\begin{equation}
T^{\mu\nu}_F=\left[\{J^\mu, J^\nu\}-g^{\mu\nu}
J^\lambda J_\lambda\right]+g^{\mu\nu}\frac{m}{2}\{\bar\psi, \psi\}
\label{t1}
\end{equation}
Thus the knowledge of the bosonized form of the currents suffices to obtain
the bosonized energy - momentum tensor in this case.

In 2+1 dimensions one can still write a formal analog of eq.(\ref{t1}),
\begin{equation}
T^{\mu\nu}_F=\frac{1}{\Lambda}\left[\{J^\mu, J^\nu\}-g^{\mu\nu}
J^\lambda J_\lambda -:\{J^\mu, J^\nu\}:+g^{\mu\nu}
:J^\lambda J_\lambda:\right]+g^{\mu\nu}\frac{m}{2}\{\bar\psi, \psi\}
\label{t}
\end{equation}
Here $:S:$ means normal ordering with respect to the perturbative vacuum.
The normal ordered term is formally of order $1/\Lambda$, and vanishes in the
continuum limit. However, being a composite operator of high dimension,
it cannot be discarded a priori.
The simplification in 1+1 dimensions
is that the normal ordered part vanishes, due to the symmetries
of the 1+1 dimensional Dirac matrices \cite{avatars}. This does
not happen in higher dimensions, and we cannot use
the Sugawara construction directly.

We can still use the construction, however,
as a guide to the form of $T^{\mu\nu}$.
The naive (i.e. neglecting the normal - ordered contributions)
Sugawara form of $T^{\mu\nu}$ contains precisely the same terms that
one gets from the explicit calculation. We will therefore take an ansatz
which is consistent
with both approaches up to finite renormalizations, and determine
the coefficients by requiring that
Lorentz invariance is recovered in the continuum limit.

The general form for the Hamiltonian density is,
\begin{equation}
T^{00}=\frac{1}{2}B^2+\frac{1}{2}E^2+
\frac{a}{e^2\Lambda}(\partial_iE_i)^2+b\frac{e^2}{4}\Lambda A^2+
\frac{1}{\Lambda}\left[c:(A\cdot\tilde E)^2:+
d:E^2A^2: \right]+fA\cdot\tilde E
\label{hamiltonian}
\end{equation}
If the theory is to be Lorentz invariant with standard transformation
properties for $E_i,B,J_0$, and $J_i$,
Maxwell's equations should be satisfied in the continuum limit. We therefore
require,
\begin{equation}
\dot B(x)=i[H,B(x)] =
-\epsilon_{ij}\partial_iE_j +O(1/\Lambda)
\end{equation}
\begin{equation}
\dot E_i=i[H,E_i]=-J_i+\epsilon_{ij}\partial_j B+O(1/\Lambda^2)
\end{equation}
Since $J_i$ itself contains terms of order $1/\Lambda$, we demand that the
second equation be satisfied to order $O(1/\Lambda^2)$.

Calculating the commutators, we find the following conditions on the
coefficients \footnote{In fact, for these values of the coefficients
the inhomogeneous Maxwell's equation is obtained exactly,
without corrections $O(1/\Lambda^{2})$. The correction term to the homogeneous
equation is of the form, $ O(1/\Lambda) = +  2/\kappa \Lambda \epsilon^{ik}
\partial_k [(A^2 E^i) + (A\tilde E)\tilde A^i - (M \kappa \Lambda /2)\tilde
A^i) ] $.
Normal - ordering the operator in the square brackets, we find that its
divergent part vanishes. Thus, even
for finite $ \Lambda $, this equation retains the form of a conservation
law.},
\begin{equation}
f=0,\ \ b=\frac{\kappa}{2}, \ \ c=-\frac{1}{\kappa}, \ \ d=\frac{1}{2\kappa}
\end{equation}
This still leaves the coefficients $\kappa$ and $a$ undetermined. Our next step
will be therefore to impose the  complete Poincar\' e algebra.

A sufficient condition for Lorentz invariance of a theory invariant
under spatial rotations and translations
is the following commutation relation \cite{schwinger},
\begin{equation}
-i[T_{00}(x),T_{00}(y)]=-(T^0_i(x)+T^0_i(y))\partial^i\delta^2(x-y)
\label{schwinger}
\end{equation}
It can be easily verified (by multiplying eq.(\ref{schwinger})
by linear functions of $x$ and $y$ and integrating) that
eq.(\ref{schwinger})
implies the closure of the Poincar\' e algebra \cite{schwinger}, with
the following genertors:
the energy $\int d^2xT^{00}(x)$, the  momentum $P_i=\int d^2xT^0_i$,
and the angular momentum and boost generators, given, respectively by,
\begin{equation}
L_i=\int d^2x [x_0T^0_i-x_iT^{00}]; \ \ L=\int d^2x \epsilon_{ij}x_iT^0_j
\label{gen}
\end{equation}
It should also be verified that the angular momentum defined in eq.(\ref{gen})
does indeed generate rotations when acting on $E_i$ and $A_i$.
Calculating the commutator in eq. (\ref{schwinger}), we find for the momentum
density,
\begin{equation}
T^0_i=B\tilde E_i-2a\kappa\partial_jE_jA_i
\label{toi}
\end{equation}
$$
+\partial_j\left[\frac{4a(q_1+q_2)+r_1+2r_2}{\kappa}\epsilon_{ij}
A\cdot\tilde E-\frac{4a(q_1+q_2)-r_1-2r_2}{\kappa}\delta_{ij}A\cdot E\right]
$$
$$
+m\frac{p_1+p_2}{2\kappa}\partial_j(A_i\tilde A_j+A_j\tilde A_i)+O(1/\Lambda)
$$
In arriving at this expression we
have extracted finite parts of the formally vanishing operators
using the operator product expansion (normal ordering) with the
UV asymptotics of the propagators given in eq.(\ref{correl}).
The relevant Wick contractions are,
\begin{eqnarray}
lim_{x-y\rightarrow 0}\frac{1}{e^2}
<\partial_iE_i(x)E_j(y)>O(y)&=&(q_1+q_2)\Lambda^2\partial_jO(y) \\ \nonumber
lim_{x-y\rightarrow 0}<\partial_iE_i(x)A_j(y)>O(y)&
=&-\frac{1}{2}(p_1+p_2)m\Lambda\epsilon_{jl}\partial_lO(y) \\ \nonumber
lim_{x-y\rightarrow 0}<\epsilon_{ki}\partial_kA_i(x)A_j(y)>O(y)&
=&-\frac{1}{2}(r_1+2r_2)\Lambda\epsilon_{jl}\partial_lO(y) \\ \nonumber
lim_{x-y\rightarrow 0}<\epsilon_{ki}\partial_kA_i(x)E_j(y)>O(y)&
=&\frac{1}{2}(p_1+p_2)m\Lambda\partial_jO(y)
\label{ope}
\end{eqnarray}

Requiring that the angular momentum generator has the standard "orbital" and
"spin" parts,
\begin{equation}
L\equiv\int d^2x
\epsilon_{ij}x_iT^0_j=\int d^2x\epsilon_{ij}x_iE_k\partial_jA_k- A\cdot\tilde E
\end{equation}
we obtain from eq.(\ref{toi}) the following relations,
\begin{equation}
a=\frac{1}{2\kappa}, \  \ \kappa=-\frac{2(q_1+q_2)}{r_1+2r_2}
\end{equation}
or, using eq.(\ref{expv}),
\begin{equation}
\kappa=\frac{\Lambda}{e^2}\frac{<E^2>}{<B^2>}
\label{kappa}
\end{equation}

The expression for the momentum density is also consistent with what
is expected  in a theory of a local
vector field $A_i$. The momentum operator $P_i=\int d^2x T^{0i}(x)$ generates
spatial translations by the usual rule,
\begin{equation}
[A_i(x),P_j]=-i\partial_iA_j(x)
\end{equation}
(and the same holds for $E_i$).

We have now determined all the coefficients in
the Hamiltonian density and the current
in terms of $\frac{<E^2>}{<B^2>}$ and $<A\cdot\tilde E>$.
Our final expressions are (we record their non - normal - ordered forms),
\begin{equation}
T^{00}=\frac{1}{2}B^2+\frac{1}{2}E^2+
\frac{1}{2e^2\kappa\Lambda}(\partial_iE_i)^2+\frac{e^2}{2}\kappa\Lambda A^2
\label{fin}
\end{equation}
$$
+\frac{1}{2\kappa\Lambda}\left[-2(A\cdot\tilde E)^2+
E^2A^2 \right]+M A\cdot\tilde E;
$$
\begin{equation}
T^{0i}= B\tilde E_i - (\partial_j E_j) A_i + \gamma
\partial_i(AE) + \frac{M}{8} \partial_j(A_i \tilde A_j + A_j \tilde A_i) +
O(\frac{1}{\Lambda});
\end{equation}
$$
J_i=-e\kappa\Lambda A_i+\frac{1}{e\kappa\Lambda}\left[2\tilde E_iA\cdot
\tilde E-A_iE^2\right]+\frac{M}{e} \tilde E_i
$$
with,
\begin{equation}
M=-\frac{2}{\kappa\Lambda}<A\cdot\tilde E>=\frac{4(p_1+p_2)}{\kappa}m
\end{equation}

\begin{equation}
\gamma = - \frac{2<B^2>}{\Lambda^3 \kappa}
\end{equation}

and $\kappa$ defined in eq.(\ref{kappa}).

We want to stress here that, although eq.(\ref{fin}) contains a parameter
$\kappa$, it does not define a one -  parameter set of theories.
The bosonized version of QED$_3$ corresponds to a unique choice
of $\kappa$, which satisfies eq.(\ref{kappa}). Unfortunately, since
eq.(\ref{fin}) defines a strongly interacting theory, we cannot determine
the numerical value of $\kappa$. This would involve the
solution of the model (at least in the UV region)
for arbitrary $\kappa$, calculating $<B^2>_\kappa$ and $<E^2>_\kappa$,
and solving the selfconsistency equation (\ref{kappa}).

We end this section with a comment.
In arriving at the form of eq.(\ref{hamiltonian}), and subsequently
eq.(\ref{fin}),
we have truncated the series for the Hamiltonian density at the order
$1/\Lambda$. This amounts to
neglecting terms of the form $\frac{1}{e^2\Lambda^3}E^4A^2$.
This is
consistent with the asymptotic form of the correlation functions
given in eq.(\ref{correl}), since the power counting based
on it tells us that we have kept all the relevant terms.
We can estimate the scale at which the irrelevant
terms become important. Assuming
perturbative scaling at low energies, a simple scaling
argument shows that these terms are unimportant relative to the terms
we kept in eq.(\ref{hamiltonian}) at energy scales
smaller than $\mu=(e^2\Lambda^2)^{1/3}$. This is again
consistent with our
previous estimate (see Section 3). At this scale all terms
become equally important,
and, once this happens, the scaling dimensions of
fields will cease to be canonical.
It is generally the case that the magnitudes
of irrelevant and relevant terms in a Hamiltonian
coincide at the UV cutoff scale. In our case
this translates into the statement that, near the scale $\Lambda$, the
electric field must have dimension $1$.
Although this argument does not constitute a proof of eq.(\ref{correl})
(and the proof cannot be given without solving the theory), it
shows the selfconsistency of our assumptions.

\section{Discussion}

In this paper we have performed the analog of Mandelstam's
construction in 2+1 dimensional
QED$_3$, that is, we have constructed the two -
 component Dirac spinor field entirely in terms
of the bosonic fields $E_i$, and
their conjugate momenta $A_i$. The fermionic bilinears are
local functions of the
bosonic variables. They satisfy a current algebra  which includes
Schwinger terms.
The bosonized
theory is Lorentz invariant in the continuum limit.
These aspects of our construction are very
similar to what one encounters in 1+1 dimensions.
There are, however, certain features which
are conceptually different, and which reflect
the differences between 1+1 - and 2+1 - dimensional physics.
We now make several comments on these issues.

Firstly, the fermionic operators eq.(\ref{psi}) anticommute only
at distances larger than the ultraviolet cutoff. In fact, the
phase in the anticommutation relations contains a power tail
of the form $(1/(\Lambda |x-y|)^s$, where $s$ is a number of
order one. This is closely related to the fact (discussed
in Sections 3 and 4), that the construction involves two UV scales
$\Lambda$ and $\mu\propto
(e^2\Lambda^2)^{1/3}$, and that the equivalence with continuum $QED_3$ holds
only below the lower scale $\mu$. This also means that, if we
were to discretize our
bosonic theory, the fermionic
operators would not have canonical local anticommutation
relations. The lattice spacing  $a$ should be identified with
the smallest distance scale required in our regularization
procedure, that is, $a\propto1/\Lambda$. The Fermi fields
would then anticommute only at distances larger than
$\frac{1}{\Lambda}=a$. This is contrary to the assumptions made in
the proof of
the Nielsen - Ninomiya theorem \cite{nielsen}, and provides a way
in which the theory avoids
the standard fermion - doubling problem.
In fact, recall, that a Hamiltonian
theory of one staggered lattice fermion field with {\it local}
anticommutation relations in 2+1 dimensions becomes in the continuum limit
a theory of four Fermi degrees of freedom. Our theory, however,
has only two fermionic degrees of freedom in the continuum limit.
 This aspect is different
from the 1+1 dimensional case, where a theory of one staggered
lattice fermion leads in the continuum limit to a theory of one
Dirac fermion. Accordingly, the anticommutation relations in 1+1 dimensions
are canonical at all distances up to the cutoff.

Secondly, we comment on the explicit appearance of the UV cutoff in our
final expressions, which is
also a novel feature. The UV cutoff never appeared explicitly
in 1+1 dimensional formulae, since there the theory of free fermions
bosonized onto a
theory of a free bosons. In higher dimensions there is however
no reason to expect that the same will happen. On the contrary,
one expects that fermions free in
the UV will be represented by strongly interacting bosons.
Indeed, the bosonic Hamiltonian eq.(\ref{fin}) contains,
apart from quadratic
terms, also a quartic interaction term with a coefficient of order
$1/\Lambda$.  Since this coefficient scales with the inverse power
of the cutoff, by naive power counting, the term is irrelevant,
However, this term is very important in determining the scaling
dimensions of various operators. For example, if we were to omit it, the
Hamiltonian would describe a (non - Lorentz - invariant)
theory of two free bosonic degrees of freedom $A_i$, with scaling dimensions
one. Both components of the electric field would then have
dimension 3/2. We know, however, that the divergence of electric
field (which is proportional to the fermion number density) must
have scaling
dimension 2. The formally irrelevant quartic interaction term is precisely
responsible for the required change of the scaling.

It is also
clear that this term cannot be treated in perturbation theory, since,
even though it appears to be irrelevant, its effect is not small.
The situation here is
very similar to the one in some strongly interacting 2+1 -
 dimensional theories defined at finite fixed points of the coupling
constant. The examples that recently have attracted much attention are the
four - Fermi theories \cite{baruch}. The four - Fermi interaction term is
perturbatively irrelevant in 2+1 dimensions. However, if the coupling
constant is defined at (or infinitesimally close to) its fixed point
value, a renormalizable and nontrivial theory is obtained. The scaling
dimensions of various operators at the fixed point are different
from their values in
the free theory. Since the coupling constant has the dimension of inverse
mass, its fixed point value is always of the form $g=\frac{k}{\Lambda}$,
where $\Lambda$ is the UV cutoff, and $k$ is a pure number.
The theory eq.(\ref{fin}) should be understood in the same sense.
It is defined at the fixed point value of the coupling of the naively
nonrenormalizable quartic interaction term, which is responsible
for the correct scaling of the fields. In this spirit eq.(\ref{kappa}),
although it was not derived here from corresponding $\beta$ - function,
should be understood as a fixed point condition.
The appearance of the positive powers of $\Lambda$ in eq.(\ref{fin})
is also natural, once one realizes that the theory is interacting. Generically,
in an interacting theory
any coupling constant with positive dimension must scale as a
power of the UV cutoff. This is the origin of the cutoff in the
$A^2$ term in eq.(\ref{fin}).
The appearance of the cutoff in the expression for the conserved current
eq.({\ref{current}) is then inevitable, since the form of the currents
is determined by the Hamiltonian.

Unfortunately, the very fact that the bosonic theory we obtained is strongly
interacting prevents us from analysing these questions quantitatively.
Approximation schemes that were
useful in analysing the four - Fermi interactions, such as the 1/N expansion,
or the $\epsilon$ - expansion, cannot be applied here,
since an extension of the
model either to large number of fermionic species, or to dimensions different
from 2+1 is not available at the moment.

Even so, there are several directions in which our approach can be extended.
The first, and most straightforward one, is
to understand  the
regularization ambiguity of $QED_3$ \cite{coste}.
The fermionic operators we have constructed solve the
Gauss' constraint. If one relaxes this condition, there are additional
possibilities \cite{inprep}.
One natural modification is to substitute for $E_i$
in eq.(\ref{vu}) the linear combination
$\Pi_i=E_i+\kappa e^2 \epsilon_{ij}A_j$. This gives
$\chi^\dagger\chi=\frac{1}{e}\partial_iE_i+\kappa e B$. ($\chi$ is the
modified Fermi field).
However, the coefficient $\kappa$  is not totally
arbitrary in this case. It is crucial for the derivation that
$\int d^2 x\partial_i\Pi_i$ has quantized eigenvalues. Otherwise, fermionic
operators will not be double valued, and the bilinears will not be local
bosonic operators (see Appendix C).
Since the magnetic flux in $QED_3$ is quantized
in units of $\frac{2\pi}{e}$, the coefficient $\kappa$ must
be given by $\frac{n}{2\pi}$, with $n$ - an integer. This
possible modification is the reflection of  the well known
regularization ambiguity in fermionic QED$_3$ \cite{coste},
which leads to the appearance of the induced Chern - Simons term
in the action or, in the Hamiltonian formalism, to the modification
of the Gauss' law constraint.
This modification of fermionic operators will also modify the
expressions for the fermionic bilinears, and the energy - momentum tensor.
The resulting bosonic theory can be obtained by the methods
used in this paper.

Thirdly, although the theory we have derived is Lorentz invariant, it
is not written explicitly in terms of Lorentz covariant fields.
It
would be desirable to make one more step, and find a formulation
in terms of a different, covariant bosonic field. We believe that
the most natural candidate for this is the magnetic vortex field.
In 2+1 dimensions it is a
scalar field \cite{npb}. It has the additional virtue that in
terms of it not only is the electric current trivially conserved,
but also the electric charge has a meaning of a topological winding
number, and is quantized classically \cite{review}.
The vortex operators appeared naturally in our construction of
the Fermi fields.
The counting of the number of degrees of freedom
suggests that this additional transformation
should be possible, since the complex vortex field
is equivalent to two real bosonic fields.
It would also  be interesting to extend the construction
to theories with more fermionic species. This generalization
may be simpler in terms of the scalar vortex variables.

Finally, we note that the fermionic operators
which we have constructed involve a {\it product}
of vortex and antivortex operators, carrying opposite magnetic fluxes.
Therefore the operators do not carry net magnetic flux, and thus are
{\it not} analogous to $2+1$  - dimensional dyons. The mechanism of their
anticommutativity is different from that discussed in \cite{wilczek}, and
subsequently extensively exploited in the analysis of Chern - Simons theories.
In fact, the approach to the problem of bosonization
presented here has a natural generalization to 3+1 dimensions.
The fermionic operators can be constructed quite easily in a
form analogous to eq.(\ref{psi}). The anticommutation relations
can be achieved by the mechanism described in the Appendix A.
While in 2+1 dimensions the fermionic operator creates a point charge
and an infinitesimally close
pair  consisting of a magnetic vortex and an antivortex
of half - integer strength,
in 3+1 dimensions it creates a point charge and an
infinitesimally small vortex loop, again of half - integer strength.
By averaging over the orientations of the loop, and by ensuring the correct
transformation properties under the
axial transformations, it is possible to construct
the four component Dirac spinor, and calculate the
bilinears.
The work along these lines is
currently in progress \cite{inprep}.

{\bf Acknowledgements.} We thank T. Jaroszewicz for discussions.
We are indebted to Y. Kluger, E. Langmann, E. Marino
and especially to T. Bhattacharya
for numerous discussions and many helpful suggestions.

\section{Appendix A. A simple mechanism for anticommutation.}

In this Appendix we give an intuitive explanation of the basic element
in the construction of the anticommuting operators $\psi_\alpha$,  and
point out its similarity to the 1+1 dimensional case.
Recall that, in 1+1 dimensional Abelian gauge theory,
the fermionic operators
can be expressed in terms of the gauge field as follows,
\begin{equation}
\psi_{1,2}(x)=
exp{\left [-ie\int_{-\infty}^{x}dyA(y)\right ]}
exp{\left [\mp i\frac
{\pi}{e}E(x)\right ]}
\label{a1}
\end{equation}
The two factors in eq.(\ref{a1}) have a very simple meaning. The first
exponential factor creates a unit charge at the point $x$,
\begin{equation}
e^{-ie\int_{-\infty}^{x}dyA(y)}
\frac{1}{e}\partial E(z)
e^{ie\int_{-\infty}^{x}dyA(y)}
=\frac{1}{e}\partial E(z)+\delta(x-z)
\end{equation}
The second exponential (up to a surface term) can be rewritten as,
\begin{equation}
V_\pm=exp{\left [\mp i
\pi\int_x^{+\infty} d^2y\frac{1}{e}\partial E(y)\right ]}
\end{equation}
It therefore  measures the total electric charge in the half space to the
right of the point $x$.
To see why two such operators anticommute, consider
$\psi_\alpha(x)\psi_\beta(y)$. Since either $x>y$ or $y>x$, only one of these
two operators creates the electric charge in the region where the other one
is measuring it. When we change the order of the fermionic
operators, the expression picks a phase according to $e^Ae^B=e^Be^Ae^{[A,B]}$.
But always, only one of the $V's$ measures the charge.
As a result the phase is always $\pm \pi$,
and the operators anticommute.

It is clear therefore that in any number of dimensions one can achieve
anticommutation relations (but {\it not} rotational invariance) from a
construction of the following kind,
\begin{equation}
\psi_\pm(x)=
exp{\left [i\int d^dye_i(x-y)A_i(y)\right ]}
exp{\left [\pm i\pi \int_M d^dy
\frac{1}{e}\partial_iE_i(y)\right ]}
\label{a2}
\end{equation}
where $e_i(x)$ is the field of a point charge ($\partial_ie_i(x)
=\delta^d(x)$), and the volume M over which the integral is performed in the
second factor is the half space ``to the right of''  the point $x$,
defined for example by $y_1>x_1$. The objects constructed in this way
anticommute precisely for the same reason as in 1+1 dimensions.

Although 2+1 dimensional fermionic operators which we discussed in this paper
where not represented in the form of eq.(\ref{a2}), the reason for their
anticommutativity is essentially the same. Forgetting momentarily
about the regularization,
and the factor ordering (which are crucial for the calculation of bilinears,
but not for the anticommutation properties of the operators at distant points),
we can write eq.(\ref{psieta}) as,
\begin{equation}
\psi_\alpha(x)\propto
exp{\left [i\int d^2ye_i(x-y)A_i(y)\right ]}
exp{\left [\mp\frac{i}{2e}\int d^2y\left (\theta(x-y)-\theta(y-x)-\pi)\right)
\partial_iE_i(y)\right ]}
\label{a3}
\end{equation}
For our present simplified discussion, we can think about the polar angle
$\theta(x)$ as a function
with a cut. The cut should
begin at the origin,
and go to spatial infinity. Let us choose the second axis
as the direction of this line. Then, for
$(x-y)_1>0$, we have
$\theta(x-y)-\theta(y-x)=-\pi$, and for  $(x-y)_1<0$ we have,
$\theta(x-y)-\theta(y-x)=\pi$. The second phase factor
in eq.(\ref{a3}) then becomes precisely the integral over the half space
$(x-y)_1>0$ of the electric charge density $\frac{1}{e}\partial_iE_i$, with
the coefficient $\pi$.

We therefore see, that the anticommutativity in our construction is achieved
by precisely the same mechanism as in 1+1 dimensions.

\section{Appendix B. The geometry of our construction.}
In this appendix we discuss the geometry of
our construction of the fermionic operators,
and point out its connection to the
formulation of Finkelstein and Rubinstein \cite{finkelstein}.

The function $\theta(x)$
is defined by the Cauchy - Riemann equation \cite{fradkin},
\begin{equation}
\epsilon_{ij}\partial_jG^{(2)}(x)=\frac{1}{2\pi}\partial_i\theta(x)
\end{equation}
Here $G^{(2)}(x-y)=-\frac{1}{4\pi}\ln(\mu^2 x^2)$.
The differential equation is solved by,
\begin{equation}
\theta(x)=\int_{C(M,x)} dy_i \epsilon_{ij}\frac{y_j}{y^2}
\end{equation}
where the curve $C(M,x)$ starts at a base point $M$, and ends at the point $x$.
It is convenient to take the base point $M$ to infinity. The function $\theta$
then depends only on the point $x$,
and the first homotopy class of $C$, i. e.,  the number
of times the curve $C$ winds around $0$. A point on
the universal cover can be parametrized by $x$, and the homotopy class
of this curve. Therefore $\theta$ is a single valued
function on the universal covering space \cite{es}.
Moreover, $e^{i\frac{\theta}{2}}$,
and therefore also $\psi_\alpha(x)$, is a single valued function on the double
cover of the plane ${\bf R}^2$.

We wish, however, to consider our theory as $QED_{3}$ on a plane, rather
than on its cover. There is a natural mathematical framework - introduced by
Finkelstein and Rubinstein - which, suitably generalized, accomodates this
interpretation of our construction.

To set the scene, we will now discuss the relevant setup in some generality.

Let $X$ be the (time $0$) configuraton space of the theory. We will assume,
that $X$ is a Banach space, noncompact in the natural topology.
Let the set $\{\hat\phi_i\}_{i = 1,...}$ comprise of quantum fields, i.e.,
the operator - valued distributions on $X$. We shall assume that there is
an action of the full Lorentz group on the fields $\hat\phi_{i}$, which
are bosonic quantities. The Fock space over $X$, ${\cal F}^{X}$, is the field
manifold of the theory. Let $Q$ denote the set of all quantum fields on $X$.
$Q$ is explicitly realized as the $L^{2}$ space of the functional Gaussian
measure, formally,
\begin{equation}
N\prod_{x\in X}\ \prod_{i}\ d\phi_{i}(x)\exp{\left[ - < \phi; K\phi>\right ]}
\end{equation}
for the positive definite kernel $K$ defined by the quadratic part of the
action. The integration is over ${\cal S}^{'}(X)$, which is the space of
Schwartz distributions on $X$.

Consider now a  multivalued functional of the fields $\hat\phi$,
say $\hat\Psi$. $\hat\Psi$ will in general depend not only on the fields
themselves, but also on continuous functions $\{\theta_{j}\}_{j = 1,...N}$,
defined on
the universal cover $\tilde X$ of the configuration space. We shall, however,
elect to work not with the cover, but with $X$ itself, accordingly,
we will take the domain of $\hat\Psi$ to be
$Q  \times\Pi^{- 1}\{(C(\tilde X)^{N}\}$, where $\Pi^{- 1}$ is the
natural projection from
the cover to the configuration space. In this approach, the ``functions''
$\theta$ are multivalued. We shall denote the domain of the functionals
$\hat\Psi$ by $\cal Q$.

Now, let $F_{\alpha}$ be a flow on $\tilde{ \cal Q}$. Any such flow induces  a
continuous 1 - parameter family of linear transformations on the functionals.
We will take here, for definiteness, the flow to be induced by rotations
about a fixed axis. Accordingly, we shall take $N = 1$, and $\theta_{1} =
\theta$, where $\theta$ is a polar variable parametrizing a 1 - dimensional
submanifold, perpendicular
to the rotation axis. The linear transformation is then exhibited as,

\begin{equation}
\hat\Psi(\hat\phi, \theta)
\rightarrow \hat\Psi(F^{- 1}_{\alpha} \hat\phi F_{\alpha}, \theta + \alpha)
\end{equation}

We shall take $\alpha \in [0, 2\pi]$, accordingly $F_{0} = {\bf 1}$,
$F_{2\pi}$ is a rotation by $2\pi$, and $F_{2\pi}\hat\phi = \hat\phi$.
However, $\hat\Psi$ is assumed to be multivalued, which implies, that
$\pi_{1}({\cal Q}) \neq \emptyset$. Since ${\cal Q}$ is a rather complicated
infinite - dimensional set, we usually cannot compute this homotopy group
directly. However, as we will show, we can relate this group to a
much simpler homotopy group. To this end, consider a given topological
sector of the theory, characterized by the asymptotic behavior of
the operators,
\begin{equation}
\hat\phi_{i}(x) \rightarrow \phi_{0i}
\end{equation}
Denote the corresponding Fock space superselection sector by
${\cal F}^X(\phi_{0})$. Now, consider a loop in $\tilde {\cal Q}$.
This can be viewed
as a mapping of the unit cell ${\bf I}^{dimX + 1} \rightarrow
{\cal F}^{X}(\phi_{0})$, with values of the fields and functions
on $\partial {\bf I}^{dimX + 1}$ specified. Hence,
\begin{equation}
\pi_{1}({\cal Q}) \approx \pi_{dimX + 1}({\cal F}^{X}(\phi _{0}))
\end{equation}
as is easily seen by considering $\hat\phi \equiv \phi_{0}$.

It is now obvious, how this construction should proceed in our formulation
of $QED_{3}$. We take $ X = {\bf R}^{2}$,
$\tilde X = \tilde{{\bf C}}/{\bf Z}_2$, and $\hat\phi = (E , A)$. Since we
are working in the gauge - fixed formulation, the corresponding Gaussian
measures are well - defined. We have explicitly construced a pair of
double - valued functionals $\hat\Psi$, which change their signs
under a
$2\pi$ rotation. We thus deduce, that $\pi_{1}({\cal Q}) \neq \emptyset$.
This implies, by the isomorphism shown above, that $\pi_{3}({\cal F}^{X}
(\phi_{0}))$ does not vanish. We conjecture, that this result
is tied to the fact, that while the
operators $E_{i}$ have continuous spectra, the operator $ \partial_i E_i$ has
a purely discrete spectrum. This implies, that a ``naive'' picture of
the resulting Fock space is certainly not correct, and this space has
nontrivial topological characteristics.

\section{Appendix C. Calculation of fermionic bilinears.}

In this appendix we give details of the calculation of fermionic bilinears.
We start with $\psi^\dagger_1\psi_1$,
\begin{equation}
\psi^\dagger_1(x)\psi_1(x)=
\frac{1}{8\pi}
\int d\hat\epsilon \left\{\left[
\psi^\dagger-1(x+\epsilon),\psi-1(x-\epsilon)\right]
,e^{ie\int_{x-\epsilon}^{x+\epsilon}dx_i
A_i^{T}}\right\}_{|\eta|, |\xi|, |\epsilon|\propto 1/\Lambda}
\end{equation}
We have to represent the product of fermionic operators as a single
exponential, expand it in powers of the regulators, and average over the
directions of the regulators.
Using the Baker - Campbell - Hausdorff
formula, $e^Ae^B=e^{A+B+\frac{1}{2}[A,B]}$,
which is valid
when $[A,B]$ is a c - number, we find,
\begin{equation}
\psi^\dagger_{1\xi}(\epsilon)\psi_{1\eta}(-\epsilon)=
e^{-\frac{i}{2}[\theta(\eta)-\theta(\xi)+\frac{1}{2}R_1(\epsilon,\xi,\eta)]}
e^{iF_1(A,E)}
\end{equation}
\begin{equation}
\psi_{1\eta}(-\epsilon)\psi^\dagger_{1\xi}(\epsilon)=
e^{-\frac{i}{2}[\theta(\eta)-\theta(\xi)-\frac{1}{2}R_1(\epsilon,\xi,\eta)]}
e^{iF_1(A,E)}
\end{equation}
Here the c - number phase $R_1$ is given by,
\begin{equation}
R_1(\epsilon,\xi,\eta)=\theta(\xi-2\epsilon)
-\theta(\xi+2\epsilon)+\theta(\eta-2\epsilon)-\theta(\eta+2\epsilon)
\end{equation}
and the operatorial part is,
\begin{equation}
F_1(A,E)=
e\int d^2yf_i(y)A_i(y)+
\frac{1}{2e}\int d^2ya^1_L(y)\partial_iE_i-
a^1_T(y)\epsilon_{ij}\partial_iE_j(y)
\end{equation}
with,
\begin{equation}
f_i(y)\equiv e_i(y+\epsilon)-e_i(y-\epsilon);
\end{equation}
$$
a^1_L(y)\equiv \theta(y-\epsilon+\xi)
-\theta(\epsilon+\xi-y)+\theta(\eta-\epsilon-y)-\theta(y+\epsilon+\eta)
$$
$$
\frac{1}{2\pi}
 a^1_T(y)\equiv G^{(2)}(y-\epsilon+\xi)-
G^{(2)}(\epsilon+\xi-y)+G^{(2)}(\eta-\epsilon-y)-G^{(2)}(y+\epsilon+\eta)
$$
We now expand the operatorial part in Taylor series. Since the
derivatives do not commute when acting on $\theta(x)$, one has to specify
their order. However, physically it is clear that we must first expand in
$\eta$ and $\xi$ at fixed $\epsilon$,
and therefore the order of derivatives is, in fact,
unambiguously determined. We use the following identities,
\begin{eqnarray}
\partial_i^x\int d^2ye_j(y-x)A_j(y)&=&-A_i^L(x) \\ \nonumber
\partial_i^x\int d^2y\theta(y-x)
\partial_jE_j(y)&=&2\pi\tilde E_i^L \\ \nonumber
\partial_i^x\int d^2yG^2(y-x)\epsilon_{jk}\partial_jE_k(y)&
=&\tilde E_i^T
\end{eqnarray}
where the superscripts $L$ and $T$ denote the
longitudinal and transverse parts of the fields, respectively.
Expanding the operatorial part up to the terms of second order, we obtain,
\begin{eqnarray}
e^{iF_1(A,E)}&=& 1+2ie\epsilon_iA_i^L-i\frac{2\pi}{e}(\eta-\xi)_i\tilde E_i-
i\frac{2\pi}{e}(\eta+\xi)_i
\epsilon_j\partial_j\tilde E_i-2e^2\epsilon_i\epsilon_jA_i^LA_j^L \\ \nonumber
&-&4\pi\epsilon_i
(\eta-\xi)_jA_i^L\tilde E_j-\frac{2\pi^2}{e^2}(\eta-\xi)_i(\eta-\xi)_j\tilde
E_i
\tilde E_j \\ \nonumber
e^{ie\int_{-\epsilon}^{\epsilon}dy_iA_I^T(y)}&=&1+2ie\epsilon_iA_i^T
-2e^2\epsilon_i\epsilon_jA_i^TA_j^T
\end{eqnarray}

The phase $R_1$ depends on the ratio $|\eta|/|\epsilon|$. It turns out,
however, that changing this ratio results in multiplying
${\rm Im}e^{\frac{iR_1}{4}}$ by a constant. Since this is the only
quantity that appears in the present calculation, its variation can always be
compensated by a suitable redefinition of $k$ in eq.(\ref{psi}).
We therefore evaluate $R_1$ for $|\epsilon|>>|\eta|$,
\begin{eqnarray}
R_1(\epsilon,\eta,\xi)
=2\pi\ &;& \ \epsilon_{ij}\epsilon_i\eta_j>0, \ \epsilon_{ij}\epsilon_i
\xi_j>0 \\ \nonumber
R_1(\epsilon,\eta,\xi)
=0 \ &;& \ \epsilon_{ij}\epsilon_i\eta_j>0, \ \epsilon_{ij}\epsilon_i
\xi_j<0 \\ \nonumber
R_1(\epsilon,\eta,\xi)
=-2\pi\ &;& \ \epsilon_{ij}\epsilon_i\eta_j<0, \ \epsilon_{ij}\epsilon_i
\xi_j<0 \\ \nonumber
  R_1(\epsilon,\eta,\xi)
=0 \ &;& \ \epsilon_{ij}\epsilon_i\eta_j<0, \ \epsilon_{ij}\epsilon_i
\xi_j>0
\end{eqnarray}

The next step is to calculate the averages, defined as,
\begin{equation}
<\alpha>\equiv
\frac
{i}{8\pi^3}\int d\hat\epsilon d\hat\eta d\hat\xi \alpha
e^{-i[\theta(\eta)-\theta(\xi)]}{\rm Im}
e^{-\frac{i}{4}R_1(\epsilon,\xi,\eta)}
\end{equation}
We find,
\begin{equation}
<1>=0;\  <\hat\eta_i\pm\hat\xi_i>=0;\ <(\hat\eta_i\pm\hat\xi_i)
(\hat\eta_j\pm\hat\xi_j)>=0; \ \
<\hat\epsilon_i>=0;\ \ <\hat\epsilon_i\hat\epsilon_j>=0
\end{equation}
$$
<(\hat\eta_i-\hat\xi_i)\hat\epsilon_j>=\frac{1}{4\pi}\delta_{ij}; \ \
<(\hat\eta_i+\hat\xi_i)\hat\epsilon_j>=\frac{i}{4\pi}\epsilon_{ij}
$$
Assembling the terms, we obtain,
\begin{equation}
\psi^\dagger_1\psi_1=
2\pi k^2\Lambda^2\left[-i<(\eta+\xi)_i\epsilon_j>\frac{1}{e}\partial_j\tilde
E_i-
<\epsilon_i(\eta_j-\xi_j)>\{A_i,\tilde E_j\}\right]
\end{equation}
$$
=\frac{k^2}{2}\Lambda^2|\epsilon||\eta|
[\frac{1}{2}\partial_iE_i-2A\cdot\tilde E]
$$
The calculation of $\psi^\dagger_2\psi_2$ proceeds in the same fashion.
The only difference is that the electric field enters all
expressions with
the opposite sign, and the averaging is done with the complex conjugate
phase factor,
\begin{equation}
<\alpha>\equiv
\frac
{i}{8\pi^3}\int d\hat\epsilon d\hat\eta d\hat\xi \alpha
e^{i[\theta(\eta)-\theta(\xi)]}{\rm Im}
e^{\frac{i}{4}R_1(\epsilon,\xi,\eta)}
\end{equation}
As a result the averages now become,
\begin{equation}
<(\hat\eta_i-\hat\xi_i)\hat\epsilon_j>=\frac{1}{4\pi}\delta_{ij}, \ \
<(\hat\eta_i+\hat\xi_i)\hat\epsilon_j>=-\frac{i}{4\pi}\epsilon_{ij}
\end{equation}
We therefore get,
\begin{equation}
\psi^\dagger_2\psi_2=
\frac{k^2}{2}\Lambda^2|\epsilon||\eta|
[\frac{1}{e}\partial_iE_i+2A\cdot\tilde E]
\end{equation}
Choosing the length of the regulators in an appropriate way,
\begin{equation}
|\epsilon||\eta|=k^{-2}\Lambda^{-2}
\end{equation}
we obtain,
\begin{equation}
\psi^\dagger\psi=\frac{1}{e}\partial_iE_i, \ \ \bar\psi\psi=-2A\cdot\tilde E
\end{equation}

We now turn to the calculation of spatial components of the current,
\begin{equation}
J_-(x)=\psi^\dagger_2(x)\psi_1(x)\equiv \frac{1}{8\pi}
\int d\hat\epsilon e^{-i\theta(\hat\epsilon)}\left\{\left[
\psi^\dagger_2(x+\epsilon),\psi_1(x-\epsilon)\right]
,e^{ie\int_{x-\epsilon}^{x+\epsilon}dx_i
A_i^{T}}\right\}_{|\eta|, |\xi|, |\epsilon|\propto 1/\Lambda}
\end{equation}
Following analogous steps, we find,
\begin{equation}
\psi^\dagger_{1\xi}(\epsilon)\psi_{2\eta}(-\epsilon)=i
e^{\frac{i}{2}[-\theta(\eta)-\theta(\xi)+\frac{1}{2}
R_2(\epsilon,\xi,\eta)]}e^{-i\pi\int d^2y \frac{1}{e}\partial_iE_i(y)}
e^{iF_2(A,E)}
\label{c1}
\end{equation}
\begin{equation}
\psi_{2\eta}(-\epsilon)\psi^\dagger_{1\xi}(\epsilon)
=-i
e^{\frac{i}{2}[-\theta(\eta)-\theta(\xi)-\frac{1}{2}
R_2(\epsilon,\xi,\eta)]}e^{-i\pi\int d^2y \frac{1}{e}\partial_iE_i(y)}
e^{iF_2(A,E)}
\label{c2}
\end{equation}
The operatorial phase is,
\begin{equation}
F_2(A,E)=ie\int d^2yf_i(y)A_i(y)-
\frac{i}{2e}\int d^2ya^2_L(y)\partial_iE_i+a^2_T(y)
\epsilon_{ij}\partial_iE_j(y)
\end{equation}
and,
\begin{equation}
a^2_L(y)\equiv \theta(y-\epsilon+\xi)-\theta(\epsilon+\xi-y)-
\theta(\eta-\epsilon-y)+\theta(y+\epsilon+\eta)
\end{equation}
$$
\frac{1}{2\pi}
a^2_T(y)\equiv G^{(2)}(y-\epsilon+\xi)-
G^{(2)}(\epsilon+\xi-y)-G^{(2)}(\eta-\epsilon-y)+
G^{(2)}(y+\epsilon+\eta)
$$
The c - number phase $R_2$ is given by,
\begin{equation}
R_2(\epsilon,\xi,\eta)=\theta(\xi-
2\epsilon)-\theta(\xi+2\epsilon)-\theta(\eta-2\epsilon)+\theta(\eta+2\epsilon)
\end{equation}
\begin{eqnarray}
R_2(\epsilon,\eta,\xi)
=0\ &;& \ \epsilon_{ij}\epsilon_i\eta_j>0, \ \epsilon_{ij}\epsilon_i
\xi_j>0 \\ \nonumber
R_2(\epsilon,\eta,\xi)
=-2\pi\ &;& \ \epsilon_{ij}\epsilon_i\eta_j>0, \ \epsilon_{ij}\epsilon_i
\xi_j<0 \\ \nonumber
R_2(\epsilon,\eta,\xi)
=0\ &;& \ \epsilon_{ij}\epsilon_i\eta_j<0, \ \epsilon_{ij}\epsilon_i
\xi_j<0 \\ \nonumber
R_2(\epsilon,\eta,\xi)
=2\pi \ &;& \ \epsilon_{ij}\epsilon_i\eta_j<0, \ \epsilon_{ij}\epsilon_i
\xi_j>0
\end{eqnarray}

The Taylor series of $F_2$ starts with a term of order zero,
\begin{equation}
-\frac{1}{e}\int d^2y [\theta(y)-\theta(-y)]\partial_iE_i(y)
\end{equation}
Again, using the Baker - Campbell - Hausdorff formula, we obtain,
\begin{equation}
e^{iF_2}=-e^{iF_2+\frac{1}{e}\int d^2y [\theta(y)-\theta(-y)]\partial_iE_i(y)}
e^{-\frac{1}{e}\int d^2y [
\theta(y)-\theta(-y)
]\partial_iE_i(y)}
\label{c4}
\end{equation}
The function $\theta(y)-\theta(-y)$ is equal either to $\pi$ or to $-\pi$.
The last operatorial factor in eq.(\ref{c4}) can therefore be written as,
\begin{equation}
exp{\left [-\frac{1}{e}\int d^2y [
\theta(y)-
\theta(-y)]\partial_iE_i(y)\right ]}=exp{\left [-i\pi(Q_+-Q_-)\right ]}
\end{equation}
where $Q_+$ is the total electric charge in the regions of space where
$\theta(y)-\theta(-y)=\pi$, and $Q_-$ the electric charge in the regions where
$\theta(y)-\theta(-y)=-\pi$. Multiplied by the factor $\exp \{-i\pi Q\}$
in eqs. (\ref{c1}-\ref{c2}), the operatorial factor becomes,
\begin{equation}
\exp\{-i2\pi Q_+\}
\label{c5}
\end{equation}
Since the electric charge is
quantized in sets which are sufficiently regular,
so that their Newtonian capacity can be defined,
this operator is,
in fact, the identity. Note that the quantization of charge
is crucial for the locality of $J_i$,
since otherwise the exponential
in eq.(\ref{c5}) would be a complicated nonlocal operator.
Now we can perform a straightforward Taylor expansion of the operatorial
phase in eqs.(\ref{c1}-\ref{c2}), and average over the directions of the
regulators.
The nonvanishing contributions are,
\begin{equation}
J_-=
-k^2\Lambda^2
2<\epsilon_i>eA_i-
\end{equation}
$$
-
k^2\Lambda^2
\left[
\frac{e}{3}
<\epsilon_i\epsilon_j\epsilon_k>
\partial_i\partial_jA_k-\frac{4e^3}{3}<\epsilon_i\epsilon_j\epsilon_k>
A_iA_jA_k\right]
$$
$$-4\pi^2k^2\Lambda^2
\left[\frac{1}{e}
<\epsilon_i(\xi_j+\eta_j)(\xi_k+\eta_k)>A_i\tilde E_j\tilde E_k
+\frac{i}{e^2}
<\epsilon_i(\xi_j-\eta_j)(\xi_k+\eta_k)>\partial_i\tilde E_j\tilde E_k\right]
$$
where the averages are now defined as,
\begin{equation}
<\alpha>\equiv\frac
{i}{8\pi^3}\int d\hat\epsilon d\hat\eta d\hat\xi \alpha
e^{-i[\theta(\eta)+\theta(\xi)+
\theta(\epsilon)]}{\rm Re}
e^{\frac{i}{4}R_2(\epsilon,\xi,\eta)}
\end{equation}
Calculation of $J_+$ proceeds in an analogous fashion. The expression
for the current obtained in this way is,
\begin{equation}
J_i\equiv\bar\psi\gamma_i\psi= e{\cal A} k^2 \Lambda^2|\epsilon| A_i+
\end{equation}
$$ek^2\Lambda^2 |\epsilon|^3\left[{\cal B}
(\partial_j\partial_jA_i+2\partial_i\partial_jA_j)
+{\cal C}e^2A_iA^2\right]+\frac{k^2}{e}\Lambda^2 |\epsilon||\eta|^2
\left[{\cal D}\tilde E_i(A\tilde E)+{\cal F} E^2A_i\right]
$$
The coefficients ${\cal A},{\cal B}, {\cal C}, {\cal D}$ and ${\cal F}$
can be calculated explicitly. We do not give their values, however, since they
are regularization - dependent. The final expression for $J_i$ is determined
in Sections 3 and 4 from the requirement of Lorentz invariance and tree - level
current algebra.

\end{document}